# *A Metamaterial-Inspired Model for Electron Waves in Bulk Semiconductors*

*Mário G. Silveirinha*[(1, 2)*] *and Nader Engheta*[(1)]

*(1) University of Pennsylvania, Department of Electrical and Systems Engineering, Philadelphia, PA, U.S.A., engheta@ee.upenn.edu*
*(2) University of Coimbra, Department of Electrical Engineering – Instituto de Telecomunicações, Portugal, mario.silveirinha@co.it.pt*

**Abstract**

Based on an analogy with electromagnetic metamaterials, we develop an effective medium description for the propagation of electron matter waves in bulk semiconductors with a zincblende structure. It is formally demonstrated that even though departing from a different starting point, our theory gives results for the energy stationary states consistent with Bastard's envelope function approximation in the long-wavelength limit. Using the proposed approach, we discuss the time evolution of a wave packet in a bulk semiconductor with a zero-gap and linear energy-momentum dispersion.

---

[*] To whom correspondence should be addressed: E-mail: mario.silveirinha@co.it.pt

# I. Introduction

Despite the fundamental differences between photons and electrons, such as mass, spin and statistics, there are many formal similarities between photonics and electronics [1-5], which ultimately result from the wave-particle duality of fundamental particles. For example, an electron is characterized by a de Broglie wavelength and thus can interfere with itself, analogous to interference phenomena in classical electromagnetism. Of significant relevance in the context of this work are the many parallelisms between the propagation of light and matter waves in periodic structures [1]. Similar to the electron motion in a semiconductor, the light propagation in a photonic crystal depends on a *band structure*, such that for some frequencies (energy levels) propagation is allowed and for others it is forbidden [1, 6, 7]. These analogies can be refined in the case of propagation in bulk materials, wherein the relevant physics is often determined by stationary states associated with a wavelength much larger than the characteristic spatial scale of importance. In the case of light, the complex interactions between radiation and polarizable matter (e.g. bound charges in atoms) result in a propagation velocity lower than in free-space, and, within the framework of macroscopic electrodynamics, this can be modeled by regarding the material as a continuous medium with an electric permittivity and magnetic permeability that differ from those of vacuum. Similarly, in case of a single electron propagating in an ionic lattice, the effect of the ionic electrostatic potential on the characteristic inertia of the electron can be described from a macroscopic point of view by assigning an effective mass $m^*$ to the electron that differs from its free rest mass $m_0$ [8], and this is instrumental in the study of the electron transport in semiconductors.

These similarities are also manifested in the formal mathematical structure of the equations used to calculate the stationary states of electronic and optical systems. For example, the time-independent Schrödinger equation for a single electron is analogous to the time-independent Helmholtz equation that describes the dynamics of a single component of the electromagnetic field in some scenarios [3, 5].

Despite these parallelisms, the theoretical frameworks typically adopted to describe wave propagation in electromagnetic media and in semiconductors are usually rather different. Macroscopic electrodynamics is based on the idea of averaging out the strong fluctuations of the microscopic electromagnetic fields in the vicinity of the polarizable particles and on the introduction of effective parameters, so that the dynamics of the wave propagation is formulated in terms of macroscopic fields that vary slowly on the scale of the microscopic unit cell [9]. These ideas also apply to the case of metamaterials, i.e. nanostructured composites synthesized by tailoring the geometry of bulk metals and dielectrics [10, 11]. The light propagation in a metamaterial relies on the introduction of mesoscopic effective parameters, which result from averaging out the fluctuations of the electromagnetic fields on a length scale determined by the period of the metamaterial, rather than by the atomic period as in natural media [12-17].

Quite differently, the computation of the electronic structure of semiconductors is typically based on perturbation methods, usually designated by $k \cdot p$ methods [18-27]. The $k \cdot p$ theory relies on the knowledge of the electronic band structure of highly-symmetric points of the Brillouin zone and on the symmetries of the associated wavefunctions ($u_{n0}$), which are used as a basis of states. This theory is also useful to model heterostructures, and in such problems the electron wavefunction is typically described by a multi-component vector, which characterizes the wavefunction in the basis $u_{n0}$. This is evidently quite different from the formalism of macroscopic electrodynamics, which does not rely on any form of "multi-component vectors" but rather on macroscopic fields that are smoothened versions of the microscopic fields. Nevertheless, it should be mentioned that the concept of a smoothened wavefunction is not strange to the semiconductor field, and the envelope-function approximation developed by G. Bastard [20, 28-29] is precisely based on such ideas. Even more generally, the concept of an envelope-function can be traced back to the pseudopotential method used in solid-state theory [8]. However, to the best of our knowledge, the connection between the envelope-function approximation and the methods of macroscopic electrodynamics was not much explored in the literature, apart from the theory formulated by Burt [24] and some cursory recent discussions [3, 5].

The main objective of the present work is precisely to demonstrate that the effective medium methods used in the context of macroscopic electrodynamics and in the theory of electromagnetic metamaterials can be extended to the case of the one-body Schrödinger equation, and in particular to the case of bulk semiconductors with a zincblende structure and associated semiconductor superlattices. The theory is based on our recent work [30], where a general effective medium theory that enables characterizing a wide range of physical systems described by a Hamiltonian was developed. Here, we show that such a formalism when applied to the case of a bulk semiconductor yields an effective Hamiltonian that within some approximations can be expressed in terms of an energy dependent effective mass and energy dependent effective potential. These two parameters are in some sense the semiconductor dual of the magnetic permeability and electric permittivity of an electromagnetic medium, respectively, as already pointed out in [3, 5]. Here these ideas are rigorously derived from "first-principles" (using as a starting point Kane's theory), and it is proven that in the long-wavelength limit our theory is closely related to the Bastard's envelope function approximation.

As shown in Refs. [3, 5], the proposed formalism enables making several interesting analogies between electromagnetic metamaterials and semiconductor superlattices. Superlattices were proposed by Esaki and Tsu more than forty years ago [31], and can be regarded as the semiconductor counterpart of electromagnetic metamaterials. In Ref. [3] we discussed how such analogies permit envisioning novel semiconductor materials with extreme anisotropy, such that the effective mass is zero along some preferred direction of motion and infinite for perpendicular directions. In Ref. [5] it was shown that electron tunneling in semiconductor heterostructures is related to light tunneling in electromagnetic metamaterials. One of the motivations of the present work is to put the findings of these previous studies into a more firm theoretical basis. In addition, we discuss the time evolution of the envelope wavefunction in zero-gap semiconductors with linear energy-momentum dispersion. Graphene is also characterized by linear energy dispersion [32, 33], however here we consider bulk materials, rather than a one-atom thick structure.

This paper is organized as follows. In section II, we briefly review the effective medium approach introduced in Ref. [30]. Then, in section III we show that for a bulk crystalline material the effective Hamiltonian resulting from the homogenization of the potential of the ionic lattice can be written in terms of the energy eigenstates. In section IV, we derive an exact formula for the effective Hamiltonian of bulk III-V and II-VI semiconductor compounds with a zincblende structure under the eight band Kane's approximation. In section V we use the effective Hamiltonian to compute the energy stationary states, and in section VI the similarities between the proposed formalism and the theory of electromagnetic metamaterials are highlighted. In section VII, the time evolution of a "macroscopic" electron wave packet in a semiconductor with a zero-bandgap is discussed. The conclusions are drawn in section VIII.

## II. Overview of the Effective Medium Approach

In Ref. [30] an effective medium approach was developed to characterize the stationary states and time evolution of systems whose dynamics is described generically by $\hat{H}\psi = i\hbar \frac{\partial}{\partial t}\psi$. In case of the one-body Schrödinger equation, $\psi$ corresponds to the wavefunction and $\hat{H}$ to the (microscopic) Hamiltonian of the system. This formalism also applies to the Maxwell's equations [30].

Our approach is based on the introduction of an effective Hamiltonian $\hat{H}_{ef}$, such that the time evolution of electronic states that are inherently "macroscopic" in the scale of the periodicity of the structure is described exactly by a modified Schrödinger equation. Specifically, we have:

$$\left(\hat{H}_{ef}\Psi\right)(\mathbf{r},t) = i\hbar \frac{\partial}{\partial t}\Psi(\mathbf{r},t) \tag{1}$$

where $\Psi(\mathbf{r},t) \equiv \{\psi(\mathbf{r},t)\}_{av}$ is the "macroscopic" wavefunction, which results from suitable spatial averaging of the exact microscopic wavefunction $\psi(\mathbf{r},t)$. The averaging operator $\{\ \}_{av}$ is such that $\{e^{i\mathbf{k}\cdot\mathbf{r}}\}_{av} = F(\mathbf{k})e^{i\mathbf{k}\cdot\mathbf{r}}$ with $F(\mathbf{k}) = 0$ for $\mathbf{k}$ outside the first

Brillouin zone (associated with the unit cell of the material or of the superlattice, depending on the structure of interest), and $F(\mathbf{k})=1$ otherwise. Thus, $\{\ \}_{av}$ corresponds to an ideal low-pass spatial filter, and $\Psi(\mathbf{r},t)$ may be regarded as a smoothened version of $\psi(\mathbf{r},t)$, with the strong spatial fluctuations on the scale of the unit cell filtered out. Notice that the result of applying $\{\ \}_{av}$ to a given function of the spatial coordinates is another function of the spatial coordinates. It was proven in Ref. [30] that the general form of $\hat{H}_{ef}$ is,

$$\left(\hat{H}_{ef}\Psi\right)_\sigma = \sum_{\sigma'}\int d^N\mathbf{r}'\int_0^t dt'\, h_{\sigma,\sigma'}(\mathbf{r}-\mathbf{r}',t-t')\Psi_{\sigma'}(\mathbf{r}',t') \qquad (2)$$

where $N$ represents the dimension of the system ($N=3$ for any bulk semiconductor or semiconductor superlattice), and $\sigma$ labels additional degrees of freedom of the electron wavefunction associated for example with the electron spin.

We say that a state is macroscopic when it remains invariant after spatial averaging $\{\psi\}_{av}=\psi$. The remarkable property of $\hat{H}_{ef}$ is that the time evolution determined by Eq. (1) of any initial macroscopic state $\Psi_{t=0}(\mathbf{r})$ is coincident with the result of applying the averaging operator to the microscopic wavefunction $\psi(\mathbf{r},t)$ obtained by solving $\hat{H}\psi=i\hbar\frac{\partial}{\partial t}\psi$, with $\hat{H}$ being the "microscopic" Hamiltonian and $\psi_{t=0}=\Psi_{t=0}$. In other words, if $\psi_{t=0}=\Psi_{t=0}$ and $\Psi_{t=0}$ is a macroscopic state, then for all the later time instants $t>0$ we have $\{\psi(\mathbf{r},t)\}_{av}=\Psi(\mathbf{r},t)$ and $\{\hat{H}\psi\}_{av}=\hat{H}_{ef}\Psi$, where the time evolution of $\psi$ is determined by the microscopic Hamiltonian $\hat{H}$, whereas the time evolution of $\Psi$ is determined by the macroscopic Hamiltonian $\hat{H}_{ef}$. This property is illustrated schematically in Fig. 1.

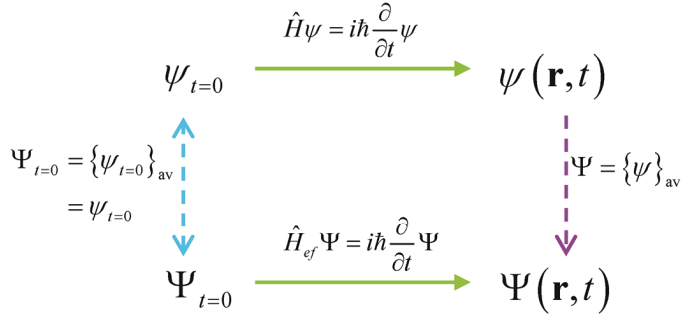

Fig. 1. (Color online) Schematic relation between the time evolutions provided by the macroscopic and microscopic Hamiltonians: for an initial macroscopic electronic state, the effective medium formulation ensures that $\Psi = \{\psi\}_{av}$ for $t > 0$.

As discussed in Ref. [30], in the Fourier domain Eq. (2) becomes a simple multiplication:

$$\left(\hat{H}_{ef}\Psi\right)_{\sigma}(\mathbf{k},\omega) = \sum_{\sigma'} h_{\sigma,\sigma'}(\mathbf{k},\omega) \Psi_{\sigma'}(\mathbf{k},\omega) \tag{3}$$

where $h_{\sigma,\sigma'}(\mathbf{k},\omega) = \int d^N \mathbf{r} \int_0^{+\infty} dt\, h_{\sigma,\sigma'}(\mathbf{r},t) e^{i\omega t} e^{-i\mathbf{k}\cdot\mathbf{r}}$ is the Fourier transform of $h_{\sigma,\sigma'}(\mathbf{r},t)$ (unilateral in time and bilateral in space), and similarly $\Psi_{\sigma}(\mathbf{k},\omega) = \int d^N \mathbf{r} \int_0^{+\infty} dt\, \Psi_{\sigma}(\mathbf{r},t) e^{i\omega t} e^{-i\mathbf{k}\cdot\mathbf{r}}$. The effective Hamiltonian is thus completely characterized by the matrix $\left[h_{\sigma,\sigma'}(\mathbf{k},\omega)\right]$, whose elements depend on the continuous parameters $\mathbf{k}$ (the wave vector) and $\omega$ (the angular frequency).

For a fixed $\mathbf{k}$ in the first Brillouin zone, one can determine $h_{\sigma,\sigma'}(\mathbf{k},\omega)$ by using the fact that for a (macroscopic) initial state such that $\psi_{\sigma}(\mathbf{r},t=0) = e^{i\mathbf{k}\cdot\mathbf{r}} \delta_{\sigma,s}$, we necessarily have $\{\psi\}_{av}(\mathbf{r},\omega) = \psi_{av} e^{i\mathbf{k}\cdot\mathbf{r}}$ and $\{(\hat{H}\psi)\}_{av}(\mathbf{r},\omega) = (\hat{H}\psi)_{av} e^{i\mathbf{k}\cdot\mathbf{r}}$ with [30]:

$$\psi_{av}(\omega) = \frac{1}{V_c} \int_\Omega d^N \mathbf{r}\, \psi(\mathbf{r},\omega) e^{-i\mathbf{k}\cdot\mathbf{r}}, \tag{4a}$$

$$\left(\hat{H}\psi\right)_{av}(\omega) = \frac{1}{V_c} \int_\Omega d^N \mathbf{r}\, \hat{H}\psi(\mathbf{r},\omega) e^{-i\mathbf{k}\cdot\mathbf{r}}. \tag{4b}$$

where $V_c$ denotes the volume of the unit cell of the direct space. Hence, making use of $\{\psi(\mathbf{r},t)\}_{av} = \Psi(\mathbf{r},t)$, $\{\hat{H}\psi\}_{av} = \hat{H}_{ef}\Psi$ and Eq. (3), it follows that:

$$\left(\hat{H}\psi\right)_{av} = \left[h_{\sigma,\sigma'}(\mathbf{k},\omega)\right]\cdot \psi_{av} \tag{5}$$

where $\psi_{av}(\omega)$ and $\left(\hat{H}\psi\right)_{av}(\omega)$ are understood as vectors. Thus, the matrix $\left[h_{\sigma,\sigma'}(\mathbf{k},\omega)\right]$ can be expressed symbolically as follows:

$$\left[h_{\sigma,\sigma'}(\mathbf{k},\omega)\right] = \left[\left(\hat{H}\psi\right)_{av}^{(s_1)};...;\left(\hat{H}\psi\right)_{av}^{(s_M)}\right]\cdot \left[\psi_{av}^{(s_1)};...;\psi_{av}^{(s_M)}\right]^{-1} \tag{6}$$

where the semicolon separates different columns of the $M \times M$ matrices, with $M$ being the number of degrees of freedom associated with $\sigma$. In the above, $\left(\hat{H}\psi\right)_{av}^{(s_i)}$ and $\psi_{av}^{(s_i)}$ are vectors defined as in Eq. (4), and the label $(s_i)$ identifies the initial macroscopic electronic state: $\psi_\sigma^{(s_i)}(\mathbf{r},t=0) \sim e^{i\mathbf{k}\cdot\mathbf{r}}\delta_{\sigma,s_i}$. Thus, in general to fully characterize $h_{\sigma,\sigma'}(\mathbf{k},\omega)$ for a fixed $\mathbf{k}$ one is required to solve $M$ microscopic problems. We also note that for $\mathbf{k}$ outside the Brillouin zone $h_{\sigma,\sigma'}(\mathbf{k},\omega) = 0$ [30].

## III. Crystalline Materials

In what follows we apply the outlined theory to electron waves in a crystalline material. The particular case of III-V and II-VI semiconductor compounds will be analyzed in details in the next section. It is assumed that the pertinent microscopic Hamiltonian in the bulk crystal for the one-electron Schrödinger equation is

$$\hat{H} = \hat{H}_0 + \frac{\hbar}{4m_0^2 c^2}(\boldsymbol{\sigma} \times \nabla V) \cdot \mathbf{p}$$
$$= \frac{p^2}{2m_0} + V(\mathbf{r}) + \frac{\hbar}{4m_0^2 c^2}(\boldsymbol{\sigma} \times \nabla V) \cdot \mathbf{p} \quad , \tag{7}$$

where $m_0$ is the free electron mass, $\mathbf{p} = -i\hbar \nabla$, $\hat{H}_0 = \frac{p^2}{2m_0} + V(\mathbf{r})$, $V(\mathbf{r})$ is the periodic crystalline potential, and the third term represents the spin-orbit interaction ($\boldsymbol{\sigma} = \sigma_x \hat{\mathbf{x}} + \sigma_y \hat{\mathbf{y}} + \sigma_z \hat{\mathbf{z}}$, with $\sigma_x$, $\sigma_y$ and $\sigma_z$ the Pauli matrices). The potential $V(\mathbf{r})$ includes both the potential from ionic lattice and some averaged potential resulting from electron-electron interactions.

To begin with, we note that the time-dependent one-body Schrödinger equation $\hat{H}\psi = i\hbar \frac{\partial}{\partial t}\psi$, with the given initial time boundary condition $\psi_{t=0}$, reduces in the Fourier domain (i.e. after a unilateral Fourier transform in the time variable: $\psi(\mathbf{r},t) \leftrightarrow \psi(\mathbf{r},\omega)$) to the time-independent equation,

$$(\hat{H} - E)\psi(\mathbf{r},\omega) = -i\hbar \psi_{t=0}(\mathbf{r}), \tag{8}$$

where we put $E = \hbar \omega$ and used the property $\frac{\partial \psi}{\partial t}(\mathbf{r},t) \leftrightarrow -i\omega \psi(\mathbf{r},\omega) - \psi_{t=0}$. From the discussion of Sect. II, it is obvious that to compute the effective Hamiltonian the initial time boundary condition should be of the form $\psi_{t=0} \sim e^{i\mathbf{k}\cdot\mathbf{r}}$. For future reference, we also note that since $\psi$ is a spinor it may be represented in vector notation as $\psi = \begin{pmatrix} \psi_\uparrow \\ \psi_\downarrow \end{pmatrix} \equiv \psi_\uparrow |\uparrow\rangle + \psi_\downarrow |\downarrow\rangle$, where the two components of the spinor are $\psi_\sigma$ with $\sigma = \uparrow, \downarrow$. As usual, $|\uparrow\rangle$ and $|\downarrow\rangle$ represent the eigenstates of the Pauli matrix $\sigma_z$.

It should be apparent that the problem of calculation of $\hat{H}_{ef}$ from the knowledge of $\hat{H}$ is a quite formidable one, of complexity comparable to the calculation of the electronic band structure of the bulk material. Next, we derive a formal expression for $\hat{H}_{ef}$ written in terms of the periodic eigenstates of the Hamiltonian $\hat{H}_0 = \frac{p^2}{2m_0} + V(\mathbf{r})$. Specifically, let us suppose that the eigenstates of $\hat{H}_0$ associated with the $\Gamma$ point of the Brillouin zone are $u_{n0}(\mathbf{r})|\sigma_n\rangle$, $n=1,2,\ldots$, where $|\sigma_n\rangle$ determines the spin state, and $u_{n0}(\mathbf{r})$ is a periodic eigenfunction. Thus, we have that

$$\hat{H}_0 u_{n0}(\mathbf{r})|\sigma_n\rangle = E_{n0} u_{n0}(\mathbf{r})|\sigma_n\rangle, \qquad \langle \sigma_m | \sigma_n \rangle \frac{1}{V_c} \int_\Omega u_m^*(\mathbf{r}) u_n(\mathbf{r}) d^3\mathbf{r} = \delta_{m,n} \qquad (9)$$

where $\Omega$ is the unit cell and $V_c$ is the corresponding volume. In order to obtain the solution of Eq. (8) for an initial time boundary condition such that $\psi_{t=0} \sim e^{i\mathbf{k}\cdot\mathbf{r}}$, we expand the microscopic wavefunction as follows

$$\psi(\mathbf{r}, \omega) = e^{i\mathbf{k}\cdot\mathbf{r}} \sum_n a_n u_{n0}(\mathbf{r})|\sigma_n\rangle, \qquad (10)$$

where $a_n$ are some unknown coefficients. It is simple to check that in these conditions, we have

$$\hat{H}\psi = e^{i\mathbf{k}\cdot\mathbf{r}} \sum_m \left( \sum_n H_{\mathbf{k},mn} a_n \right) u_{m0}(\mathbf{r})|\sigma_m\rangle \qquad (11)$$

where,

$$H_{\mathbf{k},mn} = \left(E_{n0} + \frac{\hbar^2 k^2}{2m_0}\right)\delta_{m,n} + \frac{\hbar}{m_0}\mathbf{k}\cdot\mathbf{p}_{mn}\langle\sigma_m|\sigma_n\rangle + \frac{\hbar}{4m_0^2 c^2}\langle\sigma_m|\boldsymbol{\sigma}\cdot\mathbf{q}_{mn}|\sigma_n\rangle \quad (12a)$$

$$\mathbf{p}_{mn} = \frac{1}{V_c}\int_\Omega u_m^*(\mathbf{r})\mathbf{p}\, u_n(\mathbf{r})d^3\mathbf{r} \quad (12b)$$

$$\mathbf{q}_{mn} = \frac{1}{V_c}\int_\Omega u_m^*(\mathbf{r})(\nabla V\times\mathbf{p})u_n(\mathbf{r})d^3\mathbf{r} \quad (12c)$$

To obtain this result we used Eq. (9) and, similar to the usual Kane's formalism, the k-dependent spin-orbit interaction ($\frac{\hbar}{4m_0^2 c^2}(\boldsymbol{\sigma}\times\nabla V)\cdot\hbar\mathbf{k}$) was neglected because typically it is negligibly small for the cases of interest (e.g. for binary III-V compounds [21]).

Let us suppose that the initial macroscopic state is such that $\psi_{t=0} = \frac{1}{-i\hbar}e^{i\mathbf{k}\cdot\mathbf{r}}|s\rangle$, where $|s\rangle$ determines the initial spin state. It can be shown that the effective Hamiltonian calculated below is totally independent of the normalization of the initial macroscopic state, and thus the adopted normalization is perfectly legitimate. Substituting Eq. (10) and (11) into Eq. (8), and calculating the product of both sides of the resulting equation with $\langle\sigma_m|u_{m0}^*(\mathbf{r})$ and integrating the result over the unit cell $\Omega$ of the material, we obtain a linear system of the form:

$$(H_{\mathbf{k}} - E\mathbf{1})\cdot\begin{pmatrix}a_1\\a_2\\\ldots\end{pmatrix} = \begin{pmatrix}b_1\\b_2\\\ldots\end{pmatrix} \quad (13)$$

where **1** is the identity matrix, $H_{\mathbf{k}} = [H_{\mathbf{k},mn}]$, and $b_m = \langle\sigma_m|s\rangle u_{m,av}^*$ with

$$u_{m,\mathrm{av}} = \frac{1}{V_c} \int_\Omega u_{m0}(\mathbf{r}) d^3\mathbf{r} \tag{14}$$

At this point it is convenient to denote

$$\chi_{\mathbf{k},mn} = H_{\mathbf{k},mn} - E\delta_{m,n}, \tag{15}$$

and the corresponding inverse matrix by $\left[\chi^{\mathbf{k},mn}\right] = (H_\mathbf{k} - E\mathbf{1})^{-1} = \left[\chi_{\mathbf{k},mn}\right]^{-1}$. With these notations the solution of the effective medium problem (13) can be formally written as

$$a_m = \sum_n \chi^{\mathbf{k},mn} \langle \sigma_n | s \rangle u^*_{n,\mathrm{av}}. \tag{16}$$

In the remainder of this section, we use this result to obtain an explicit formula for the effective Hamiltonian. From Eqs. (10) and (11) it is evident that $\psi_{\mathrm{av}}(\omega)$ and $\left(\hat{H}\psi\right)_{\mathrm{av}}(\omega)$, defined as in Eq. (4), are given by,

$$\psi_{\mathrm{av}}(\omega) = \sum_n a_n u_{n,\mathrm{av}} |\sigma_n\rangle, \tag{17a}$$

$$\left(\hat{H}\psi\right)_{\mathrm{av}}(\omega) = \sum_m \left(\sum_n H_{\mathbf{k},mn} a_n\right) u_{m,\mathrm{av}} |\sigma_m\rangle, \tag{17b}$$

where $u_{m,\mathrm{av}}$ satisfies Eq. (14). On the other hand, in agreement with Eq. (6), the matrix that characterizes the effective Hamiltonian can be written as:

$$H_{ef}(\mathbf{k},\omega) \equiv \left[h_{\sigma,\sigma'}(\mathbf{k},\omega)\right] = \left[\left(\hat{H}\psi\right)_{\mathrm{av}}^{(\uparrow)}; \left(\hat{H}\psi\right)_{\mathrm{av}}^{(\downarrow)}\right] \cdot \left[\psi_{\mathrm{av}}^{(\uparrow)}; \psi_{\mathrm{av}}^{(\downarrow)}\right]^{-1} \tag{18}$$

where the semicolon separates different columns of the $2\times 2$ matrices, and $\left(\hat{H}\psi\right)_{\mathrm{av}}^{(\uparrow)}$ and $\psi_{\mathrm{av}}^{(\uparrow)}$ are defined as in Eq. (17) for the case where the initial state is $\psi_{t=0} = \frac{1}{-i\hbar} e^{i\mathbf{k}\cdot\mathbf{r}} |s\rangle$, with $|s\rangle = |\uparrow\rangle$, and $(\hat{H}\psi)_{av}^{(\downarrow)}$ and $\psi_{av}^{(\downarrow)}$ are defined similarly. It is useful to note that from Eqs. (13) and (17),

$$(\hat{H}\psi)_{av}(\omega) = E\psi_{av}(\omega) + \sum_m \langle \sigma_m|s\rangle |u_{m,av}|^2 |\sigma_m\rangle. \tag{19}$$

Substituting this result into Eq. (18), it is seen that the effective Hamiltonian can be written as:

$$H_{ef}(\mathbf{k},\omega) = E\mathbf{1} + \left[ \sum_m \langle \sigma_m|\uparrow\rangle |u_{m,av}|^2 |\sigma_m\rangle ; \sum_m \langle \sigma_m|\downarrow\rangle |u_{m,av}|^2 |\sigma_m\rangle \right] \cdot \left[ \psi_{av}^{(\uparrow)} ; \psi_{av}^{(\downarrow)} \right]^{-1} \tag{20}$$

With the help of Eq. (17) each element of the matrices can be written explicitly as shown below:

$$H_{ef}(\mathbf{k},\omega) = E\mathbf{1} + \begin{bmatrix} \sum_m |\langle \sigma_m|\uparrow\rangle|^2 |u_{m,av}|^2 & \sum_m \langle \sigma_m|\downarrow\rangle\langle\uparrow|\sigma_m\rangle |u_{m,av}|^2 \\ \sum_m \langle \sigma_m|\uparrow\rangle\langle\downarrow|\sigma_m\rangle |u_{m,av}|^2 & \sum_m |\langle \sigma_m|\downarrow\rangle|^2 |u_{m,av}|^2 \end{bmatrix} \times$$

$$\begin{bmatrix} \sum_m a_m^{(\uparrow)} u_{m,av} \langle\uparrow|\sigma_m\rangle & \sum_m a_m^{(\downarrow)} u_{m,av} \langle\uparrow|\sigma_m\rangle \\ \sum_m a_m^{(\uparrow)} u_{m,av} \langle\downarrow|\sigma_m\rangle & \sum_m a_m^{(\downarrow)} u_{m,av} \langle\downarrow|\sigma_m\rangle \end{bmatrix}^{-1} \tag{21}$$

where $a_n^{(\uparrow)}$ is the solution of (13) when $|s\rangle = |\uparrow\rangle$, and $a_n^{(\downarrow)}$ is defined similarly. Using now Eq. (16) we obtain the following formula for the effective Hamiltonian relative to the basis $|\uparrow\rangle$ and $|\downarrow\rangle$:

$$H_{ef}(\mathbf{k},\omega) = E\mathbf{1} + \begin{bmatrix} \sum_m |\langle\sigma_m|\uparrow\rangle|^2 |u_{m,\mathrm{av}}|^2 & \sum_m \langle\sigma_m|\downarrow\rangle\langle\uparrow|\sigma_m\rangle |u_{m,\mathrm{av}}|^2 \\ \sum_m \langle\sigma_m|\uparrow\rangle\langle\downarrow|\sigma_m\rangle |u_{m,\mathrm{av}}|^2 & \sum_m |\langle\sigma_m|\downarrow\rangle|^2 |u_{m,\mathrm{av}}|^2 \end{bmatrix} \times$$

$$\begin{bmatrix} \sum_{m,n} \chi^{\mathbf{k},mn} u^*_{n,\mathrm{av}} u_{m,\mathrm{av}} \langle\sigma_n|\uparrow\rangle\langle\uparrow|\sigma_m\rangle & \sum_{m,n} \chi^{\mathbf{k},mn} u^*_{n,\mathrm{av}} u_{m,\mathrm{av}} \langle\sigma_n|\downarrow\rangle\langle\uparrow|\sigma_m\rangle \\ \sum_{m,n} \chi^{\mathbf{k},mn} u^*_{n,\mathrm{av}} u_{m,\mathrm{av}} \langle\sigma_n|\uparrow\rangle\langle\downarrow|\sigma_m\rangle & \sum_{m,n} \chi^{\mathbf{k},mn} u^*_{n,\mathrm{av}} u_{m,\mathrm{av}} \langle\sigma_n|\downarrow\rangle\langle\downarrow|\sigma_m\rangle \end{bmatrix}^{-1}$$

(22)

The above result is exact in case we consider a complete set of periodic eigenstates of $\hat{H}_0$ ($u_{n0}(\mathbf{r})|\sigma_n\rangle$ $n=1,2,\ldots$). It should be noted that in the spectral domain the effective Hamiltonian is represented by a $2\times 2$ matrix. In the next section, we obtain an explicit approximate analytical formula for $H_{ef}$ for the case of III-V and II-VI binary compounds.

## IV. Bulk III-V and II-VI compounds

Next, we consider the particular case of bulk III-V and II-VI semiconductors with a zincblende structure. The zincblende lattice consists of face-centered cubic lattice with two atoms per elementary cell, and is characteristic of binary III-V compounds such as GaAs, GaSb, InSb, and II-VI compounds such as HgTe and CdTe [18, 20]. The exact effective Hamiltonian is written in terms of the periodic eigenfunctions of $\hat{H}_0 = \frac{p^2}{2m_0} + V(\mathbf{r})$ as in Eq. (22). However, such Bloch functions are seldom known explicitly. To make some progress, we need to introduce some simplifying assumptions. Specifically, in the same spirit of $k \cdot p$ theory, we suppose that in the *energy range* of interest the envelope of $\psi$ can be written as linear combination of a few energy-eigenfunctions of the Hamiltonian $\hat{H}_0 = \frac{p^2}{2m_0} + V(\mathbf{r})$. For semiconductors with the zincblende structure there are typically eight relevant crystal states of $\hat{H}_0$ for energies in the range determined by the valence and conduction bands [18, 20, 21, 22]. Each state is doubly degenerate because $\hat{H}_0$ does not depend explicitly on the electron spin. The relevant states are labeled as $|S\sigma\rangle$, $|X\sigma\rangle$, $|Y\sigma\rangle$ and $|Z\sigma\rangle$, with $\sigma=\uparrow,\downarrow$, and the associated wavefunctions have the symmetries of the atomic $s, x, y, z$ functions under the operations of the tetrahedral group. The states $|S\sigma\rangle$ are associated with the edge of the conduction band ($E_{n0} = E_{s0}$), whereas $|X\sigma\rangle$, $|Y\sigma\rangle$ and $|Z\sigma\rangle$ are all degenerate at the $\Gamma$ point and are associated with the edge of the valence bands ($E_{n0} = E_{p0}$). Thus, in what follows we evaluate Eq. (22) restricting the summation to the contributions of the above-mentioned eight crystal states. Notice that within this approximation the matrix $H_{\mathbf{k}} = \left[ H_{\mathbf{k},mn} \right]$ [Eq. 12] can be identified with the usual Hamiltonian matrix used in the context of Kane's approach [18, 20, 21, 22] [actually, in the $k \cdot p$ theory typically the adopted basis of functions is not exactly the one described above, but rather another equivalent basis whose elements are linear combinations of $|S\sigma\rangle$, $|X\sigma\rangle$, $|Y\sigma\rangle$ and $|Z\sigma\rangle$; obviously, the effective Hamiltonian is independent of the considered basis].

Let us then consider the above-mentioned basis of expansion functions, so that $n, m = 1, 2, ..., 8$ in Eq. (22). The first important observation is that because of the symmetries of $|X\sigma\rangle$, $|Y\sigma\rangle$ and $|Z\sigma\rangle$, it is evident that:

$$u_{m,\text{av}} = 0, \quad \text{for states of the form } |X\sigma\rangle, |Y\sigma\rangle \text{ and } |Z\sigma\rangle. \tag{23}$$

On the other hand, for states of the form $|S\sigma\rangle$, we have $u_{m,\text{av}} \equiv u_{s,\text{av}}$, where $u_{s,\text{av}}$ is some constant. Substituting these results into Eq. (22), and assuming that the eigenfunctions $|S\uparrow\rangle$, $|X\downarrow\rangle$, $|Y\downarrow\rangle$ and $|Z\uparrow\rangle$ are associated with the indices $m=1,2,3,4$, respectively, and that $|S\downarrow\rangle$, $|X\uparrow\rangle$, $|Y\uparrow\rangle$ and $|Z\downarrow\rangle$ are associated with the indices $m=5,6,7,8$, respectively, it is readily found that:

$$H_{ef}(\mathbf{k},\omega) = E\mathbf{1} + |u_{s,av}|^2 \begin{bmatrix} \chi^{\mathbf{k},11}|u_{s,av}|^2 & \chi^{\mathbf{k},15}|u_{s,av}|^2 \\ \chi^{\mathbf{k},51}|u_{s,av}|^2 & \chi^{\mathbf{k},55}|u_{s,av}|^2 \end{bmatrix}^{-1}$$
$$= E\mathbf{1} + \begin{bmatrix} \chi^{\mathbf{k},11} & \chi^{\mathbf{k},15} \\ \chi^{\mathbf{k},51} & \chi^{\mathbf{k},55} \end{bmatrix}^{-1} \quad (24)$$

Therefore, $H_{ef}(\mathbf{k},\omega)$ is independent of $u_{s,av}$ and is written exclusively in terms of a few elements of the matrix $\left[\chi^{\mathbf{k},mn}\right] = (H_{\mathbf{k}} - E\mathbf{1})^{-1} = \left[\chi_{\mathbf{k},mn}\right]^{-1}$.

To determine the required elements of $\left[\chi^{\mathbf{k},mn}\right]$, first we will evaluate explicitly $H_{\mathbf{k}}$. To this end, we note that because of the symmetries of band edge functions the coefficients $\mathbf{p}_{mn}$, given by Eq. (12b), vanish except if one of the indices is associated with a valence band state and another with a conduction band state. Specifically, we have $\langle m|\mathbf{p}|m\rangle = 0$ with $m=S,X,Y,Z$, $\langle m|\mathbf{p}|n\rangle = 0$ with $m,n=X,Y,Z$, and $\langle S|\mathbf{p}|X\rangle = \langle S|p_x|X\rangle \hat{\mathbf{x}}$, etc, with

$$P = -i\frac{\hbar}{m_0}\langle S|p_x|X\rangle = -i\frac{\hbar}{m_0}\langle S|p_y|Y\rangle = -i\frac{\hbar}{m_0}\langle S|p_z|Z\rangle \quad (25)$$

being $P$ Kane's parameter [18, 20, 21, 22]. To see which elements of $\mathbf{q}_{mn} = \langle m|\nabla V \times \mathbf{p}|n\rangle$ [Eq. 12c] are different from zero, we start by noting that the z-component of this vector is $(\mathbf{q}_{mn})_z = -i\hbar\left\langle m\left|\frac{\partial V}{\partial x}\frac{\partial}{\partial y} - \frac{\partial V}{\partial y}\frac{\partial}{\partial x}\right|n\right\rangle$. Therefore, the operator $\nabla V \times \mathbf{p}$ changes both the parity of $y$ and $x$ of $|n\rangle$ and leaves the parity of $z$ unchanged.

Thus, the only values of $m,n=S,X,Y,Z$ for which $(\mathbf{q}_{mn})_z$ is different from zero are clearly ($m=X$ and $n=Y$) or ($m=Y$ and $n=X$). In conclusion, this discussion shows that:

$$\frac{3i\hbar}{4m_0^2 c^2}(\mathbf{q}_{mn})_l = \varepsilon_{lmn}\Delta, \qquad m,n,l=X,Y,Z \tag{26}$$

where $\varepsilon_{lmn}$ is the Levi-Civita symbol, and $\Delta = \frac{3i\hbar}{4m_0^2 c^2}\left\langle X \left| \frac{\partial U}{\partial x}p_y - \frac{\partial U}{\partial y}p_x \right| Y \right\rangle$ is the spin-orbit split-off energy [18, 20, 21, 22]. Finally, we note that $(\mathbf{q}_{mn})_l = 0$ if either $m$ or $n$ are equal to $S$. Based on Eqs. (25)-(26), it is possible to write $H_\mathbf{k} = [H_{\mathbf{k},mn}]$ [Eq. 12a] as follows [$P$ and $\Delta$ can be assumed real-valued]:

$$[H_{\mathbf{k},mn}] = \begin{bmatrix} E'_{s0} & 0 & 0 & iPk_z & 0 & iPk_x & iPk_y & 0 \\ 0 & E'_{p0} & -\frac{\Delta}{3i} & -\frac{\Delta}{3} & -iPk_x & 0 & 0 & 0 \\ 0 & +\frac{\Delta}{3i} & E'_{p0} & \frac{\Delta}{3i} & -iPk_y & 0 & 0 & 0 \\ -iPk_z & -\frac{\Delta}{3} & -\frac{\Delta}{3i} & E'_{p0} & 0 & 0 & 0 & 0 \\ 0 & iPk_x & iPk_y & 0 & E'_{s0} & 0 & 0 & iPk_z \\ -iPk_x & 0 & 0 & 0 & 0 & E'_{p0} & \frac{\Delta}{3i} & \frac{\Delta}{3} \\ -iPk_y & 0 & 0 & 0 & 0 & -\frac{\Delta}{3i} & E'_{p0} & \frac{\Delta}{3i} \\ 0 & 0 & 0 & 0 & -iPk_z & \frac{\Delta}{3} & -\frac{\Delta}{3i} & E'_{p0} \end{bmatrix} \tag{27}$$

where we put $E'_{s0} = E_{s0} + \frac{\hbar^2}{2m_0}k^2$, $E'_{p0} = E_{p0} + \frac{\hbar^2}{2m_0}k^2$, with $E_{s0}$ being the energy eigenvalue associated with the conduction band of $\hat{H}_0$, and $E_{p0}$ being the energy eigenvalue associated with the valence bands of $\hat{H}_0$. Again, we emphasize that the above matrix differs from the standard Hamiltonian matrix used in the $k \cdot p$ approach, simply because we are considering the basis $|S\uparrow\rangle$, $|X\downarrow\rangle$, $|Y\downarrow\rangle$, $|Z\uparrow\rangle$, $|S\downarrow\rangle$, $|X\uparrow\rangle$, $|Y\uparrow\rangle$ and $|Z\downarrow\rangle$.

Straightforward calculations show that $\left[\chi^{k,mn}\right] = (H_k - E\mathbf{1})^{-1}$ is such that $\chi^{k,15} = \chi^{k,51} = 0$ and $\chi^{k,11} = \chi^{k,55}$. Hence, from Eq. (24) it follows that somewhat surprisingly, despite the spin-orbit interaction is considered, the effective Hamiltonian reduces to a scalar $H_{ef}(\mathbf{k},\omega) = E + 1/\chi^{k,11}$. Moreover, by inverting $H_k - E\mathbf{1}$, we obtain the following explicit formula for the effective Hamiltonian,

$$H_{ef}(\mathbf{k},\omega) = E + \frac{-\frac{2}{9}\Delta^2 \tilde{E}_s + \left(\tilde{E}_p - \frac{\Delta}{3}\right)\left(\tilde{E}_s \tilde{E}_p - k^2 P^2\right)}{\left(\tilde{E}_p - \frac{2}{3}\Delta\right)\left(\tilde{E}_p + \frac{\Delta}{3}\right)} \tag{28}$$

where we defined $\tilde{E}_s = E_{s0} - E + \frac{\hbar^2}{2m_0}k^2$ and $\tilde{E}_p = E_{p0} - E + \frac{\hbar^2}{2m_0}k^2$, with $E = \hbar\omega$. Therefore, within the same approximations that are usually considered in the $k \cdot p$ approach our effective medium theory predicts that the dynamics of envelope of the electron wavefunction can be described by a scalar effective Hamiltonian, whose formula in the spectral domain is given by Eq. (28). This is the main result of this section.

It is convenient to introduce the notations $\varepsilon_P = 2P^2 m_0 / \hbar^2$, $E_v = E_{p0} + \frac{\Delta}{3}$, and $E_c = E_{s0}$, so that the effective Hamiltonian becomes after some simplifications:

$$H_{ef}(\mathbf{k},\omega) = E + \tilde{E}_c + \frac{\hbar^2}{2}k^2 \frac{\varepsilon_P}{3m_0}\left(\frac{-2}{\tilde{E}_v} + \frac{-1}{\tilde{E}_v - \Delta}\right) \tag{29}$$

where $\tilde{E}_c = E_c - E + \frac{\hbar^2}{2m_0}k^2$ and $\tilde{E}_v = E_v - E + \frac{\hbar^2}{2m_0}k^2$. As is well-known (and will also be discussed in the next section) $E_c$ and $E_v$ determine the energy levels at the edges of the conduction and light-hole bands, respectively. It is also interesting to point out that in case the spin-orbit coupling is neglected ($\Delta = 0$) one has:

$$H_{ef}(\mathbf{k},\omega) = E + \tilde{E}_c - \frac{\hbar^2}{2}k^2 \frac{\varepsilon_P}{m_0} \frac{1}{\tilde{E}_v}, \qquad \text{for } \Delta = 0. \qquad (30)$$

## V. Stationary states

The stationary states of the energy operator can be readily obtained using the effective medium Hamiltonian. Indeed, it was proven in Ref. [30] that the eigenvalues $E$ of the microscopic $\hat{H}$ Hamiltonian are the same as the eigenvalues of the exact effective Hamiltonian $\hat{H}_{ef}$ (there can however exist some isolated exceptions that are discussed below). Within the approximations made in the previous section, we can restate this property as: the energy eigenvalues $E$ computed using the standard $k \cdot p$ approach based on a multi-component wavefunction are the same as the eigenvalues of the effective Hamiltonian $H_{ef}$ given by Eq. (29).

To show this more explicitly, we note that for a wavefunction $\Psi$ with a time variation of the form $e^{-i\frac{E}{\hbar}t}$ (an energy eigenstate) and space variation of the form $e^{i\mathbf{k}\cdot\mathbf{r}}$, the homogenized Schrödinger's equation (1) reduces to:

$$\left(H_{ef}(\mathbf{k},E) - E\right) \cdot \Psi = 0 \qquad (31)$$

where for simplicity we replaced $E = \hbar\omega$ in the argument of $H_{ef}$, so that it is regarded a function of energy. Hence, since in our case $H_{ef}$ is a scalar, from Eq. (29) we obtain the following characteristic equation for the energy eigenstates:

$$\tilde{E}_c + \frac{\hbar^2}{2}k^2 \frac{\varepsilon_P}{3m_0}\left(\frac{-2}{\tilde{E}_v} + \frac{-1}{\tilde{E}_v - \Delta}\right) = 0. \qquad (32)$$

It should be noted that the eigenfunctions associated with the eigen-energies are doubly degenerate and are of the form $\Psi = \left[\Psi_\uparrow |\uparrow\rangle + \Psi_\downarrow |\downarrow\rangle\right] e^{i\mathbf{k}\cdot\mathbf{r}}$, being $\Psi_\sigma$ some constants. These eigenfunctions are coincident with the spatially averaged eigenfunctions determined using the standard $k \cdot p$ approach, because it is evident from the previous section [see Eq. (23)] that $\{|p\sigma\rangle\}_{av} = 0$ for $p$=X,Y,Z and $\sigma = \uparrow,\downarrow$ [21].

Defining the bandgap energy $E_g = E_c - E_v$, equation (32) can be recast into the form:

$$E'(E' - E_g)(E' + \Delta) - k^2 P^2 \left(E' + \frac{2\Delta}{3}\right) = 0, \qquad E' = E - E_v - \frac{\hbar^2}{2m_0}k^2 \qquad (33)$$

This is coincident with the secular equation derived originally by E. Kane [21], which describes the dispersion of the conduction, light-hole and split-off band. For small values of $k$ the corresponding solutions are [21]:

$$E = E_v + E_g + \frac{\hbar^2}{2m_0}k^2 + \frac{P^2\left(E_g + \frac{2\Delta}{3}\right)}{E_g(E_g + \Delta)}k^2 \qquad \text{(conduction band)} \qquad (34a)$$

$$E = E_v + \frac{\hbar^2}{2m_0}k^2 - \frac{2}{3}\frac{P^2}{E_g}k^2 \qquad \text{(light-hole band)} \qquad (34b)$$

$$E = E_v - \Delta + \frac{\hbar^2}{2m_0}k^2 - \frac{1}{3}\frac{P^2}{(\Delta + E_g)}k^2 \qquad \text{(split-off band)} \qquad (34c)$$

It is interesting to note that our formalism does not predict the heavy-hole band that is obtained (even though with a physically incorrect dispersion) within the $k \cdot p$ approach approximation. Since we mentioned before that our formalism should predict the same eigenvalues as the microscopic Hamiltonian, an explanation is in need. The reason, as discussed in Ref. [30], is that strictly speaking such a property only holds for eigenstates $|n\rangle$ whose projection into the subspace of macroscopic eigenstates is different from zero. Specifically, if we denote $\hat{O}_{av}$ the operator corresponding to the operation of spatial averaging (see Ref. [30]), and if $\hat{H}|n\rangle = E_n|n\rangle$ then it is only possible to guarantee that $E_n$ is as well an eigenvalue of the effective Hamiltonian if $\hat{O}_{av}|n\rangle \neq 0$.

This readily explains why this theory does not predict the heavy-hole band. Indeed, within the approximations of Sect. IV (i.e. within the approximations of the $k \cdot p$ approach) the eigenstates associated with the heavy-hole band are linear combinations of kets of the form $|p\sigma\rangle$ with $p$=X,Y,Z and $\sigma = \uparrow, \downarrow$, and as discussed previously all these states are projected by $\hat{O}_{av}$ into the null state. The states for which $\hat{O}_{av}|n\rangle = 0$ can be regarded as "dark" states (borrowing a term conventionally used in photonics), in the sense that they cannot be excited if the initial electron state is macroscopic. It should however be noted that in case one considers an extended set of expansion functions (rather than just eight as in Sect. IV) to include the effect of remote bands, in principle the heavy-hole band should be predicted by our formalism because in principle, at least for $k \neq 0$, it is not expected that $\hat{O}_{av}|n\rangle = 0$ for the states associated with heavy-hole band.

## VI. Local Effective Parameters and Analogy with Electromagnetic Metamaterials

The effective Hamiltonian $H_{ef}$, given by Eq. (29), depends on the wave vector $\mathbf{k} \leftrightarrow -i\nabla_{\mathbf{r}}$ in a relatively complicated manner. Hence it is not straightforward to obtain a formulation of the problem in the space domain, and even if we invert the Fourier transform of $H_{ef}$ with respect to the wave vector the resulting expression may be too complicated to allow for further progress. In this respect, the situation is quite analogous to the case of electromagnetic metamaterials, which in general must be described using a dielectric function of the form $\overline{\overline{\varepsilon}}_{ef}(\omega, \mathbf{k})$ to account for the effects of spatial dispersion [14, 15, 34]. However, when the spatial dispersion effects are weak, it is possible to characterize the mesoscopic response of the material using *local effective parameters*, i.e. effective parameters that are $\mathbf{k}$ independent [14, 15, 34]. The knowledge of the local effective parameters (if they can be defined) is of paramount importance when one is interested in the study of wave phenomena in the vicinity of an interface between two bulk materials, wherein the formulation of boundary conditions is crucial.

For electromagnetic metamaterials (as well as for natural media) characterized by weak spatial dispersion, the nonlocal dielectric function can be approximated by [14, 15, 35]:

$$\frac{\overline{\overline{\varepsilon}}_{ef}}{\varepsilon_0}(\omega, \mathbf{k}) \approx \overline{\overline{\varepsilon}}_r + c^2 \frac{\mathbf{k}}{\omega} \times \left( \overline{\overline{\mu}}_r^{-1} - \overline{\overline{\mathbf{I}}} \right) \times \frac{\mathbf{k}}{\omega} \tag{35}$$

where $\bar{\bar{\mathbf{I}}}$ is the identity dyadic, and $\bar{\bar{\varepsilon}}_r$ and $\bar{\bar{\mu}}_r$ are the *local* effective permittivity and permeability tensors of the material, respectively. In particular, it follows from the above expression that the local effective permittivity is related to the nonlocal dielectric function as [14],

$$\bar{\bar{\varepsilon}}_r(\omega) = \frac{1}{\varepsilon_0} \bar{\bar{\varepsilon}}_{ef}(\omega, \mathbf{k} = 0), \tag{36a}$$

whereas the *zz*-component of the magnetic permeability can be written in terms of the derivatives of the nonlocal dielectric function with respect to the wave vector (evidently, it is possible to write similar formulas for the remaining components of the permeability tensor) [14]

$$\frac{\mu_{zz}}{\mu_0} \equiv \mu_{r,zz}(\omega) = \frac{1}{1 - \left(\frac{\omega}{c}\right)^2 \frac{1}{2\varepsilon_0} \left.\frac{\partial^2 \varepsilon_{ef,yy}}{\partial k_x^2}\right|_{\mathbf{k}=0}}. \tag{36b}$$

Thus, in case of weak spatial dispersion, it is possible to determine the local effective parameters directly from the spatially dispersive dielectric function [14].

Can these ideas be adapted to the case of electron waves? The generalization is straightforward: indeed, typically the relevant physical phenomena in the II-VI and III-V binary compounds considered in this work are mainly determined by the form of the electronic structure in the vicinity of the $\Gamma$ point. Thus, to study such phenomena it is enough to consider small values of $\mathbf{k}$, and this idea is actually already implicit in the approximations made in Sect. IV. Hence, we can expand the effective Hamiltonian in a Taylor series in powers of $\mathbf{k}$. Since $H_{ef}$ is an even function of $\mathbf{k}$, it follows that:

$$H_{ef}(\mathbf{k}, E) \approx H_{ef}(\mathbf{k}=0, E) + \frac{1}{2}\sum_{i,j}\frac{\partial^2 H_{ef}}{\partial k_i \partial k_j}k_i k_j \,. \tag{37}$$

If we introduce an effective potential $V_{ef}$ such that

$$V_{ef}(E) = H_{ef}(\mathbf{k}=0, E), \tag{38}$$

and an effective mass tensor $\overline{\overline{M}}_{ef}$ such that

$$\overline{\overline{M}}_{ef}^{-1}(E) = \frac{1}{\hbar^2}\left[\frac{\partial^2 H_{ef}}{\partial k_i \partial k_j}\bigg|_{\mathbf{k}=0}\right], \tag{39}$$

the effective Hamiltonian may be rewritten as,

$$H_{ef}(\mathbf{k}, E) \approx \frac{\hbar^2}{2}\mathbf{k}\cdot\overline{\overline{M}}_{ef}^{-1}\cdot\mathbf{k} + V_{ef}(E) \tag{40}$$

which justifies the used nomenclature. Comparing the above formulas with (36) it should be clear that $V_{ef}(E)$ plays a role similar to $\overline{\overline{\varepsilon}}_r$ in the electromagnetic problem, whereas $\overline{\overline{M}}_{ef}$ plays a role similar to $\overline{\overline{\mu}}_r$. This will be further elaborated shortly.

The effective parameters $V_{ef}$ and $\overline{\overline{M}}_{ef}$ are by definition independent of the wave vector, and hence are local parameters. However, they may depend on the energy $E = \hbar\omega$. Based on Eq. (40), we can readily invert the Fourier transform with respect to the wave vector in Eq. (3), to find that in the space domain:

$$(\hat{H}_{ef}\Psi)(\mathbf{r}, E) = -\frac{\hbar^2}{2}\nabla\cdot\left[\overline{\overline{M}}_{ef}^{-1}\cdot\nabla\Psi(\mathbf{r}, E)\right] + V_{ef}(E)\Psi(\mathbf{r}, E) \tag{41}$$

In particular, the energy eigenstates are solutions of the following time-independent Schrödinger-type equation $\left(\hat{H}_{ef}\Psi\right)(\mathbf{r},E) - E\Psi(\mathbf{r},E) = 0$, or equivalently:

$$-\frac{\hbar^2}{2}\nabla\cdot\left[\overline{\overline{M}}_{ef}^{-1}\cdot\nabla\Psi\right] + V_{ef}(E)\Psi = E\Psi.  \qquad (42)$$

Let us now discuss what this theory gives us for the particular case wherein the Hamiltonian is described by Eq. (29). Straightforward calculations show that the effective potential is a constant (independent of energy)

$$V_{ef}(E) = E_c, \qquad (43)$$

whereas the effective mass is a scalar such that,

$$\frac{1}{M_{ef}} = \frac{1}{m_0} + v_P^2\left(\frac{2}{E-E_v} + \frac{1}{E-E_v+\Delta}\right) \qquad (44)$$

where we defined $v_P = \sqrt{\varepsilon_P/(3m_0)}$, which has dimensions of velocity.

We remind that $E_v = E_{\Gamma_8}$ is the (light-hole) valence band edge energy, $E_c = E_{\Gamma_6}$ is the conduction band edge energy. Hence, within the approximation implicit in Eq. (37) the dynamics of the electron wavefunction can be described simply in terms of an energy-dependent effective mass and in terms of an effective potential. Curiously, formula (44) is well known in the context of Bastard's envelope function approximation [20, p. 88]. Here, we rediscovered the result of Bastard based on the effective medium approach proposed in our earlier work [30]. We believe that this analysis puts into a more firm stand the actual physical meaning of $M_{ef}$ as a component of an effective Hamiltonian with the properties described in Sect. II, and also highlights the analogies with the homogenization of electromagnetic metamaterials [14, 15].

It is interesting to point out that in case of narrow gap semiconductors $\Delta$ is typically at least a few times larger than $|E_g|$. Thus, for such materials and for energies in the interval defined by $E_c$ and $E_v$ only the first term inside brackets in Eq. (44) is relevant. In these conditions the dispersive effective mass may be approximated by

$$M_{ef} \approx \frac{E - E_v}{2v_P^2}. \tag{45}$$

To further explore the similarities between the electronic and photonic problems, next we outline an elementary correspondence between the solutions of the Schrödinger and Maxwell's equations, which agrees with what has already been reported in [3, 5]. To this end, we consider the propagation of electromagnetic waves in a two-dimensional structure whose permeability $\overline{\overline{\mu}}_r = \mu_{xx}\hat{\mathbf{x}}\hat{\mathbf{x}} + \mu_{yy}\hat{\mathbf{y}}\hat{\mathbf{y}} + \mu_{zz}\hat{\mathbf{z}}\hat{\mathbf{z}}$ and permittivity $\varepsilon_r$ are independent of $y$. Furthermore, we assume that the electromagnetic fields do not vary with $y$ and that the electric field is polarized along $y$: $\mathbf{E} = E_y \hat{\mathbf{y}}$. In such conditions, the electromagnetic field is completely characterized by $E_y$, which satisfies:

$$\nabla \cdot \left[ \left( \frac{1}{\mu_{zz}} \hat{\mathbf{x}}\hat{\mathbf{x}} + \frac{1}{\mu_{xx}} \hat{\mathbf{z}}\hat{\mathbf{z}} \right) \cdot \nabla E_y \right] + \frac{\omega^2}{c^2} \varepsilon_r E_y = 0. \tag{46}$$

Comparing this formula with the time-independent Schrödinger equation (42), it is possible to make the following correspondences (for a fixed frequency $\omega$): $\psi \leftrightarrow E_y$, $\sqrt{2}\frac{1}{\hbar} \leftrightarrow \frac{\omega}{c}$, and most importantly,

$$E - V_{ef} \leftrightarrow \varepsilon_r$$
$$\overline{\overline{M}}_{ef}^{-1} \leftrightarrow -\hat{\mathbf{y}} \times \overline{\overline{\mu}}_r^{-1} \times \hat{\mathbf{y}} \quad . \tag{47}$$

Obviously, we do not attribute any physical meaning to the above correspondences, and regard them only as a tool to transform solutions of one of the problems into solutions of the other problem. The correspondences can be useful to better understand wave phenomena, and suggest that for such purposes $E - V_{ef}$ may be regarded as the dual of the electric permittivity, whereas $\overline{\overline{M}}_{ef}^{-1}$ can be regarded as the dual of $-\hat{\mathbf{y}} \times \overline{\overline{\mu}}_r^{-1} \times \hat{\mathbf{y}}$. In the isotropic case, we can simply state that $M_{ef}$ is the dual of $\mu_r$. Evidently, this type of analogies is not new, and similar (and sometimes equivalent) ideas were considered in other works [4, 5].

To illustrate the typical dependence of $E - V_{ef}$ and $M_{ef}$ with the energy, we show in Fig. 2 the effective parameters of several semiconductor binary compounds with a zincblende structure. Without loss of generality, $E_v = E_{\Gamma_8}$ is taken equal to zero in these plots. The points where $E = E_v$ and $E = E_c$ are marked with dashed vertical gridlines in the Fig. 2, and represent the edges of the light-hole valence and conduction bands, respectively. As seen, at the edge of the light-hole valence band the dispersive effective mass, $M_{ef}$, crosses zero, whereas the parameter $E - V_{ef}$ crosses zero at the edge of the conduction band. It is important to stress that $M_{ef} = M_{ef}(E)$ is totally different from the effective mass $M^* = \left[ \frac{1}{\hbar^2} \frac{\partial^2 E}{\partial k_i \partial k_j} \right]^{-1}$ calculated from the curvature of the energy dispersion, i.e. from Eqs. (34). In particular, at the edge of the valence band $M_{ef}$ crosses zero, but $M^*$ is obviously different from zero. It is also interesting to mention that $M_{ef}$ also crosses zero at $E = E_v - \Delta$ (see Fig. 2), which corresponds to the edge of the split-off valence band.

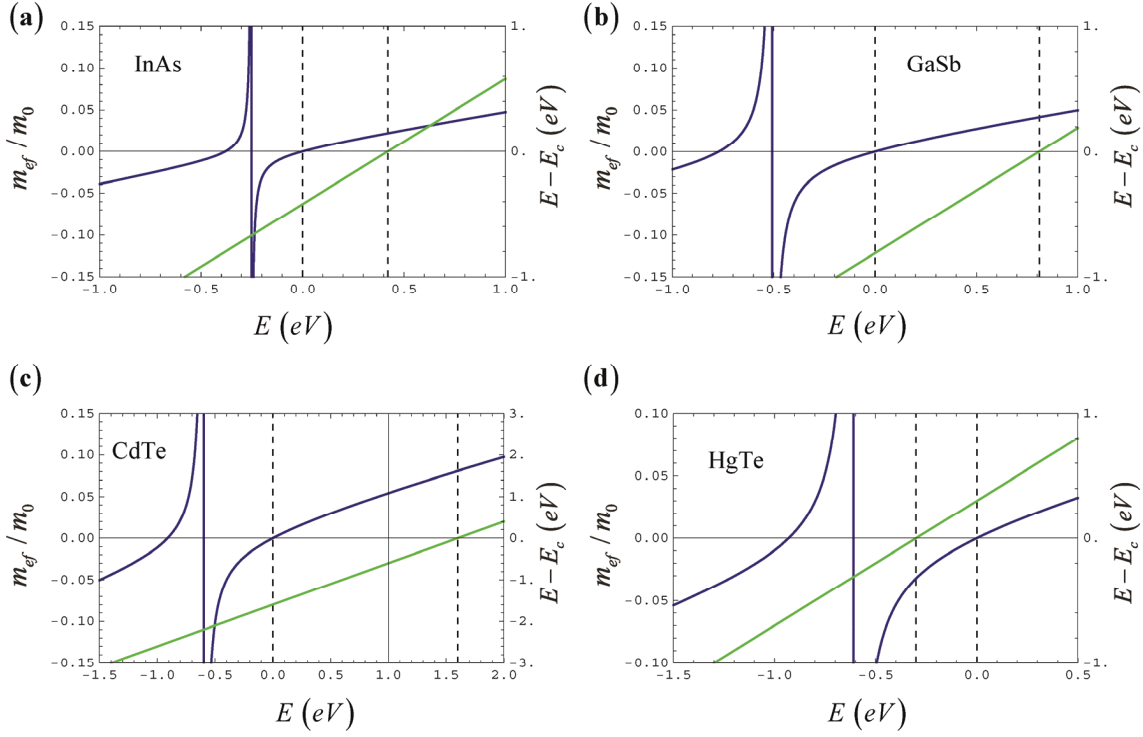

Fig. 2. (Color online) Effective parameters of several bulk semiconductors as a function of the energy $E$. The energy level of the edge of the valence band is taken arbitrarily equal to zero in all cases ($E_v = 0$). Blue (dark gray) lines [associated with left-hand side scale of the plots]: $m_{ef}/m_0$; Green (light gray) lines [associated with right-hand side scale of the plots]: $E - E_c$ in [eV]. The dashed vertical gridlines indicate the edges of the light-hole valence and conduction bands, and delimit a bandgap. The parameters of the semiconductors are taken from Refs. [36, 37].

For the binary compounds InAs, GaSb and CdTe the edge of the conduction band lies above the edge of the light-hole valence band. This corresponds to a positive bandgap $E_g = E_c - E_v > 0$ which is the typical situation in most semiconductors. For these materials $M_{ef} > 0$ and $E - V_{ef} < 0$ in the bandgap, and thus in such energy range the semiconductor has a behavior similar to a metamaterial with $\mu > 0$ and $\varepsilon < 0$ (ENG material) [3, 5]. In the valence and conduction bands the signs of $M_{ef}$ and $E - V_{ef}$ are the same: in the conduction band the semiconductor has a behavior analogous to a metamaterial with $\mu > 0$ and $\varepsilon > 0$ (DPS material), whereas in the valence band it is analogous to a material with $\mu < 0$ and $\varepsilon < 0$ (DNG material).

On the other hand, the semimetal HgTe has the unusual property that $E_g = E_c - E_v < 0$, i.e. it has a negative bandgap energy, so that the band with *s*-type symmetry lies *below* the bands *p*-type symmetry. Due to this inverted band structure, in Fig. 2d the edge of the valence band (where $M_{ef} = 0$) lies above the edge of the conduction band. In the bandgap, we have $M_{ef} < 0$ and $E - V_{ef} > 0$, and thus this material may behave similar to a metamaterial with $\mu < 0$ and $\varepsilon > 0$ (MNG material) [3, 5]. For energy levels immediately below the lower edge of the bandgap both effective parameters are simultaneously negative, whereas for energy levels above the upper edge of the bandgap both effective parameters are simultaneously positive.

In the framework of the model based on the parameters $V_{ef}$ and $M_{ef}(E)$, the energy stationary states can be determined by solving Eq. (42), which for the case of Bloch-modes in a bulk material (associated with the quasi-momentum **k**) reduces to:

$$\frac{\hbar^2}{2} \frac{k^2}{M_{ef}} + V_{ef} = E. \qquad (48)$$

In Fig. 3 we depict the electronic band structure calculated by solving the above equation with respect to $E$ for HgTe and for the ternary alloy Hg$_{0.75}$Cd$_{0.25}$Te. The bandgap energy $E_g = E_{g,x}$ of the ternary compound Hg$_{1-x}$Cd$_x$Te is calculated using Hansen's formula at zero temperature [37, 38], where $x$ represents the mole fraction. Thus, the effective potential of each material can be written as $V_{ef,x} = E_{v,x} + E_{g,x}$. On the other hand, the valence band offset $\Lambda(x) = E_{v,x=0} - E_{v,x}$ between Hg$_{1-x}$Cd$_x$Te and HgTe, can be estimated to vary with the mole fraction as $\Lambda(x) = 0.35x\ [eV]$ [39], so that $E_{v,x} = E_{v,H_gT_e} - \Lambda(x)$. The value of $E_{v,H_gT_e}$ can be arbitrarily chosen, and fixes the reference energy. In this work, $E_{v,H_gT_e}$ is set equal to zero. The velocity $v_P = \sqrt{\varepsilon_P/(3m_0)}$ can be estimated equal to $v_P = 1.06 \times 10^6\ m/s$ and the spin-orbit split-off energy as $\Delta = 0.93\ [eV]$. These two parameters are to a first approximation independent of the mole fraction [37].

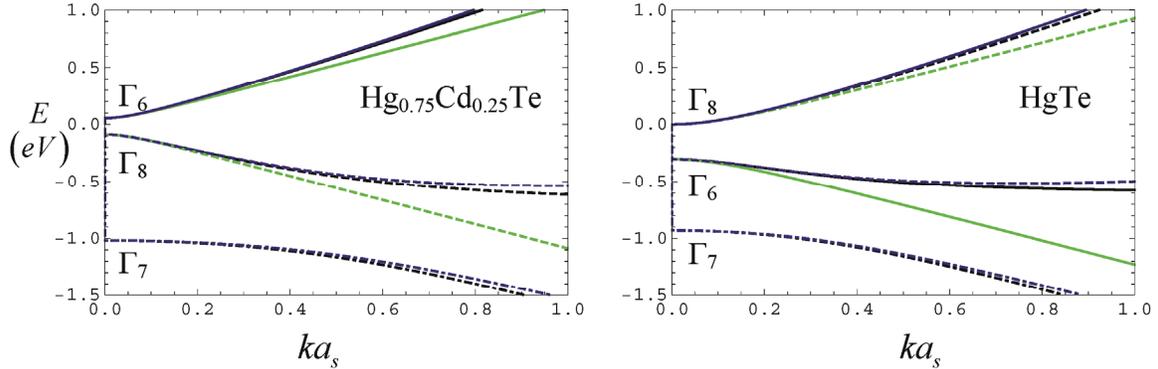

Fig. 3. (Color online) Electronic band structure of Hg$_{0.75}$Cd$_{0.25}$Te (left-panel) and HgTe (right-panel), calculated with the envelope-function approximation. The quasi-momentum $k$ is normalized to atomic lattice constant $a_s = 0.65nm$ [37]. Black lines: calculated with $M_{ef}$ defined as in Eq. (44); Blue (dark gray) lines: Result obtained with standard k·p theory [Eq. (27)]; Green (light gray) lines: linear dispersive mass approximation [Eq. (45)]. The linear dispersive mass approximation does not predict the spin-orbit split-off ($\Gamma_7$) band. The heavy-hole band is not shown.

The black lines in Fig. 3 represent the solution of (48), using the "exact" formula for $M_{ef}$ given by Eq. (44), which is practically coincident with the result obtained with standard k·p theory (blue lines). As expected, in agreement with Sect. V, three bands are found: a conduction band (solid curves), a light-hole valence band (dashed curves) and a split-off spin-orbit valence band (dot-dashed curves). On the other hand, the green curves in Fig. 3 represent the solution of (48) obtained with the linear dispersive mass approximation [Eq. (45)]. Because this approximation assumes that the spin-orbit split-off energy is very large only the conduction and light-hole valence bands are predicted. Thus, the linear dispersive mass approximation is effectively a two-band model. As seen, this approximation can be quite accurate for low-energy excitations with **k** close to the Γ point.

It is interesting to further discuss the properties of the electronic structure, as predicted by the linear dispersive mass approximation. For $M_{ef} = \frac{E - E_v}{2v_P^2} = \frac{E - V_{ef} + E_g}{2v_P^2}$, being $E_g = E_c - E_v = V_{ef} - E_v$ the bandgap energy, equation (48) is equivalent to

$$(\hbar k v_P)^2 = (E - V_{ef})(E - V_{ef} + E_g), \tag{49}$$

which is evidently a quadratic function of $E$. Solving with respect to $E$ it is found that:

$$E - V_{ef} = -\frac{E_g}{2} \pm \sqrt{\left(\frac{E_g}{2}\right)^2 + (\hbar k v_P)^2} . \tag{50}$$

For $\hbar k v_P \ll |E_g|/2$ one can write $E \approx V_{ef} + \left(-\frac{E_g}{2} \pm \left|\frac{E_g}{2}\right|\right) \pm \frac{(\hbar k v_P)^2}{|E_g|}$, and this yields the following approximate dispersions for the conduction and light-hole bands

$$E|_c \approx V_{ef} + \frac{(\hbar k v_P)^2}{E_g}, \qquad E|_{lh} \approx V_{ef} - E_g - \frac{(\hbar k v_P)^2}{E_g}, \qquad (51)$$

respectively. Therefore, within the indicated approximations the valence and conduction bands are exactly symmetric with respect to the center of the gap, being each the mirror of the other. The standard ("group") effective mass for the conduction band, $m_c^* \equiv \left[\frac{1}{\hbar^2}\frac{\partial^2 E|_c}{\partial k_i \partial k_j}\right]^{-1}$, and for the valence band, $m_{lh}^* \equiv -\left[\frac{1}{\hbar^2}\frac{\partial^2 E|_{lh}}{\partial k_i \partial k_j}\right]^{-1}$, satisfy:

$$m_c^* = m_{lh}^* = \frac{E_g}{2v_P^2} \qquad (52)$$

This result is consistent with fact that for narrow-gap semiconductors $m_c^* \approx m_{lh}^*$ [36]. Notice that $m_c^*$ and $m_{lh}^*$ are negative when the bandgap energy is negative (e.g. for HgTe [36]). Evidently, the energy dispersion in Eq. (51) can also be obtained directly from Eqs (34a)-(34b) by neglecting the small term $\frac{\hbar^2}{2m_0}k^2$ and considering the limit $\Delta/|E_g| \to \infty$.

The previous discussion assumes implicitly that the bandgap energy is different from zero. If $E_g = 0$ it is found from Eq. (50) that:

$$|E - V_{ef}| = \hbar k v_P. \qquad (53)$$

In these conditions, the relation between energy and quasi-momentum becomes linear. This is similar to graphene [33], except that here we have a bulk three-dimensional semiconductor. Notice that $v_P$ plays a role analogous to the Fermi velocity in graphene [33]. The possibility of the emergence of a zero-gap in bulk semiconductor alloys and in semiconductor superlattices has been discussed in many works [40-45].

In particular, a zero-bandgap may occur for a specific value of the mole fraction of the ternary alloy Hg$_{1-x}$Cd$_x$Te [38, 43]. According to Hansen's formula [37, 38], the bandgap energy should vanish at $x \approx 0.17$ in the zero-temperature limit (right-hand side panel of Fig. 4). In this case, the "group" effective mass vanishes, $m_c^* = m_{lh}^* = 0$, and the electronic transport may be mainly determined by the velocity $v_P = 1.06 \times 10^6 \, m/s$. In the left-panel of Fig. 4 we depict the electronic band structure of Hg$_{1-x}$Cd$_x$Te for different values of the mole fraction. The results were computed using Eq. (48) with the linear dispersive mass approximation. In agreement with the previous discussion, for $x \approx 0.17$ the energy-momentum relation becomes linear, which is consistent with the fact that the mobility of HgCdTe compounds may be remarkably high [37, 42, 43].

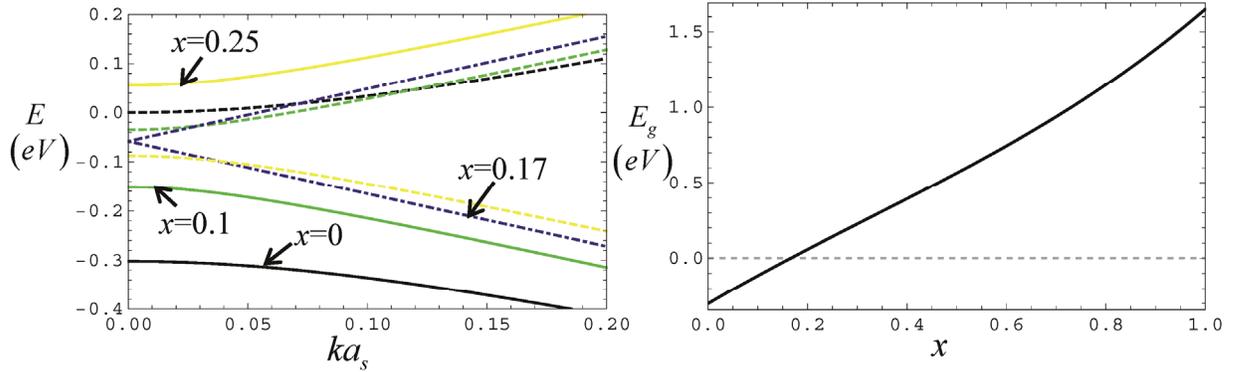

Fig. 4. (Color online) Left-panel: Electronic band structure of Hg$_{1-x}$Cd$_x$Te for different values of the mole fraction. The quasi-momentum $k$ is normalized to atomic lattice constant $a_s = 0.65 nm$ [37]. The solid curves represent the conduction band ($\Gamma_6$), whereas the dashed lines represent the light-hole valence band ($\Gamma_8$). For $x \approx 0.17$ (dot-dashed blue curves) the two bands are in contact and $m_c^* = m_{lh}^* = 0$. Right-panel: Bandgap energy for Hg$_{1-x}$Cd$_x$Te as a function of the mole fraction, following Hansen's formula [37, 38].

# VII. Time Evolution of a Macroscopic Wave Packet in a Zero-Gap Structure

The previous sections were mostly focused in the electronic band structure. However, as discussed in Sect. II and in Ref. [30], our theory can also be applied to study the dynamics in time of the wavefunction, and ensures that for initial macroscopic states, the effective medium theory describes exactly the time evolution of the envelope of the microscopic wavefunction (Fig. 1).

To further highlight these properties, next we consider a hypothetical bulk semiconductor with a zero-gap. As mentioned previously, a ternary alloy of Hg$_{1-x}$Cd$_x$Te with $x \approx 0.17$ may be an example. In the zero-gap case, using the linear dispersive mass approximation [Eq. (45)], the effective Hamiltonian (40) can be written explicitly as:

$$\hat{H}_{ef}(\mathbf{r}, E) = -\frac{\hbar^2 v_P^2}{E - V_{ef}} \nabla^2 + V_{ef} \tag{54}$$

Thus, using the Fourier transform pair $\dfrac{1}{E - V_{ef}} \leftrightarrow \dfrac{1}{i\hbar} e^{-i\frac{V_{ef}}{\hbar}t} u(t)$ where $u(t)$ is the Heaviside step function, it is possible to write in the time domain that [compare with Eq. (2)]:

$$\left(\hat{H}_{ef} \Psi\right) = i\hbar v_P^2 e^{-i\frac{V_{ef}}{\hbar}t} \int_0^t dt' \, e^{i\frac{V_{ef}}{\hbar}t'} \nabla^2 \Psi(\mathbf{r}, t') + V_{ef} \Psi(\mathbf{r}, t), \qquad t > 0. \tag{55}$$

Strictly speaking the above manipulations are only possible if Eq. (54) is valid for arbitrary $E$, whereas in practice we know that it is only valid for energies such that $E \sim V_{ef}$. Nevertheless, our approximation may be acceptable if the Fermi level of the system is close to $V_{ef}$.

Within the considered model, the effective medium Schrödinger equation (1) in the time domain becomes:

$$i\hbar \frac{\partial \Psi}{\partial t}(\mathbf{r},t) = i\hbar v_P^2 e^{-i\frac{V_{ef}}{\hbar}t} \int_0^t dt' \, e^{i\frac{V_{ef}}{\hbar}t'} \nabla^2 \Psi(\mathbf{r},t') + V_{ef} \Psi(\mathbf{r},t), \quad t > 0. \tag{56}$$

This equation can be solved univocally for a given initial time boundary condition $\Psi_{t=0}(\mathbf{r})$. Defining $\phi(\mathbf{r},t) = e^{i\frac{V_{ef}}{\hbar}t} \Psi(\mathbf{r},t)$ it is straightforward to verify that $\phi$ satisfies,

$$\frac{\partial \phi}{\partial t} = v_P^2 \int_0^t dt' \, \nabla^2 \phi(\mathbf{r},t'), \quad t > 0. \tag{57}$$

Differentiating both sides of the equation with respect to time, one sees that $\phi(\mathbf{r},t)$ satisfies the wave equation

$$\frac{1}{v_P^2} \frac{\partial^2 \phi}{\partial t^2} - \nabla^2 \phi = 0. \tag{58}$$

This was in part expected because of the assumed linear energy-momentum relation in the spectral domain [Eq. (53)]. Thus, we can obtain the time evolution of the system by simply solving the wave equation. The initial time boundary conditions are $\phi(\mathbf{r}, t=0) = \Psi_{t=0}(\mathbf{r})$ and $\frac{\partial \phi}{\partial t}(\mathbf{r},0) = \frac{\partial}{\partial t}\left[e^{i\frac{V_{ef}}{\hbar}t}\Psi(\mathbf{r},t)\right] = e^{i\frac{V_{ef}}{\hbar}t} \frac{\partial}{\partial t}\left[i\frac{V_{ef}}{\hbar}\Psi(\mathbf{r},0) + \frac{\partial \Psi}{\partial t}(\mathbf{r},0)\right]$. But from Eq. (56) it follows that $\frac{\partial \Psi}{\partial t}(\mathbf{r}, t=0) = -i\frac{V_{ef}}{\hbar}\Psi(\mathbf{r}, t=0)$, and hence we conclude that:

$$\phi(\mathbf{r},t=0) = \Psi_{t=0}(\mathbf{r}) \quad \text{and} \quad \frac{\partial \phi}{\partial t}(\mathbf{r},t=0) = 0 \tag{59}$$

are the initial time boundary conditions. Notice that even though the wave equation is of second order in time, there is only one non-trivial initial time boundary condition, consistent with the fact that the effective medium Schrödinger equation (56) is a differential equation of first order in time.

To see the implications of these findings, we consider a simple one-dimensional problem such that the wavefunction is independent of the coordinates $y$ and $z$ and $\Psi_{t=0}(\mathbf{r}) = \Psi_{t=0}(x)$, where $\Psi_{t=0}(x)$ is a given complex function. The solution of the wave equation subject to the initial time boundary conditions (59) can be integrated explicitly and is:

$$\phi(x,t) = \frac{1}{2}\left[\Psi_{t=0}(x - v_P t) + \Psi_{t=0}(x + v_P t)\right]. \tag{60}$$

In particular, one sees that if the initial electronic macroscopic state is localized in space, let us say close to the origin, it will be split into two wave packets that propagate exactly with velocity $v_P$ along the positive and negative $x$-axis, respectively. This is evidently very different from the dynamics of a free-electron subject to the standard Schrödinger equation, because in the zero-gap system for sufficiently large $t$ the wave packet is effectively split into two, as if we had two particles moving in opposite directions. This result may appear paradoxical but it is actually a direct consequence of the fact that *in the time domain* our theory only applies to *macroscopic* states. As proven next, in a zero-gap semiconductor a macroscopic state is such that the probability of the electron energy being in the conduction band is exactly the same as probability of the electron energy being in the valence band.

To prove this, let us consider first that the initial state is $\psi_{t=0}(x) = e^{ik \cdot x}$, where $k$ is arbitrarily fixed. As before, the *microscopic* wavefunction is denoted by $\psi$. Evidently, $\psi_{t=0}$ can be written as a linear combination of the energy Bloch eigenstates of the *microscopic* Hamiltonian associated with the wave vector $k$:

$$e^{ik \cdot x} = \sum_n c_n \psi_{n\mathbf{k}}(\mathbf{r}). \tag{61}$$

In the above, $\psi_{n\mathbf{k}}(\mathbf{r})$ are the microscopic energy eigenstates and $c_n$ are the coefficients of the expansion. Within our two-band model, there are only two such microscopic states: one associated with the conduction band ($\psi_{c,k}(x)$) and another with the valence band ($\psi_{v,k}(x)$). Thus, within the microscopic theory the initial state should be regarded as a superposition of the states in the conduction and valence bands: $\psi_{t=0} = c_E \psi_{c,k}(x) + c_H \psi_{v,k}(x)$. It is shown in the Appendix that for a bulk semiconductor with a zero-gap the coefficients $c_E$ and $c_H$ are such that $c_E = -c_H$. Hence, we have

$$e^{ik \cdot x} \approx c_E \left[ \psi_{c,k}(x) - \psi_{v,k}(x) \right]. \tag{62}$$

Thus, the macroscopic state $\psi_{t=0}$ is a superposition with *equal weights* of eigenstates of the conduction and valence bands. Note that $\psi_{c,k}(x)$ and $\psi_{v,k}(x)$, i.e. the microscopic eigenstates associated with the conduction and valence bands, in general are not macroscopic states (i.e. $\psi \neq \{\psi\}_{av}$).

Next, we consider a general localized initial macroscopic state. This state is necessarily a superposition of plane waves (with $k$ in the first Brillouin zone), and hence from Eq. (62) it is of the form:

$$\psi_{t=0}(x) \approx \int dk\, c_k \left[\psi_{c,k}(x) - \psi_{v,k}(x)\right]. \tag{63}$$

This formula confirms that the probability of the electron being in the conduction band is exactly the same as that of being in the valence band, and in essence this is why a macroscopic localized electron wave packet splits into two. Indeed, each plane wave of the wave packet is a sum of conduction (forward electron wave propagating with velocity $+v_P$) and valence (backward electron wave propagating with velocity $-v_P$) eigenstates [see Eq. (62)], such that the probability of the electron being in either the conduction or valence band is the same. For completeness, next we re-derive Eq. (60) directly from the microscopic theory.

To do this first we note that the result of averaging a Bloch mode is simply a plane wave [30]. Hence, it is possible to write,

$$\{\psi_{i,k}(x)\} = \Psi_{i,k} e^{ikx}, \quad i=c,v \tag{64}$$

for some constants $\Psi_{i,k}$. Because the initial state is macroscopic $\psi_{t=0}(x) = \{\psi_{t=0}(x)\}_{av} \equiv \Psi_{t=0}(x)$, it follows that

$$\psi_{t=0}(x) \approx \int dk\, c_k \left(\Psi_{c,k} - \Psi_{v,k}\right) e^{ikx}. \tag{65}$$

This formula will be useful later. From Eq. (63) it is clear that the wavefunction as a function of time is given by:

$$\psi(x,t) \approx \int dk\, c_k \left(\psi_{c,k}(x) e^{-i\frac{E_{c,k}}{\hbar}t} - \psi_{v,k}(x) e^{-i\frac{E_{v,k}}{\hbar}t}\right). \tag{66}$$

where $E_{c,k}$ and $E_{v,k}$ are the energy dispersions of the eigenstates. In particular, the spatially averaged wavefunction $\Psi(x,t) = \{\psi(x,t)\}_{av}$ satisfies:

$$\Psi(x,t) \approx \int dk\, c_k \left( \Psi_{c,k} e^{-i\frac{E_{c,k}}{\hbar}t} - \Psi_{v,k} e^{-i\frac{E_{v,k}}{\hbar}t} \right) e^{ikx}. \tag{67}$$

We prove in the Appendix, that for the considered basis $\Psi_{c,k} = -\Psi_{v,k}$. Hence, using

$$E_{c,k} - V_{ef} = -E_{v,k} + V_{ef} = \hbar |k| v_P \quad [\text{Eq. (53)}] \quad \text{and}$$

$$\left( e^{-i\frac{E_{c,k}}{\hbar}t} + e^{-i\frac{E_{v,k}}{\hbar}t} \right) e^{ikx} = e^{-i\frac{V_{ef}}{\hbar}t} \left( e^{ik(x-v_P t)} + e^{ik(x+v_P t)} \right), \text{ it is found that:}$$

$$\begin{aligned}
\Psi(x,t) &= e^{-i\frac{V_{ef}}{\hbar}t} \int dk\, c_k \Psi_{c,k} \left( e^{ik(x-v_P t)} + e^{ik(x+v_P t)} \right) \\
&= e^{-i\frac{V_{ef}}{\hbar}t} \int dk\, c_k \left( \Psi_{c,k} - \Psi_{v,k} \right) \frac{1}{2} \left( e^{ik(x-v_P t)} + e^{ik(x+v_P t)} \right). \\
&= \frac{1}{2} e^{-i\frac{V_{ef}}{\hbar}t} \left[ \psi_{t=0}(x - v_P t) + \psi_{t=0}(x + v_P t) \right]
\end{aligned} \tag{68}$$

This completely agrees with the result obtained using the effective medium theory [Eq. (60)], as we wanted to prove.

From a more fundamental perspective, one may also say that the form the solution (60) is in some sense a consequence of the uncertainty relations, which are preserved by the effective Hamiltonian description. Indeed, a solution of the form $\phi(x,t) = \Psi_{t=0}(x - v_P t)$ would allow for both the localization in space and for the knowledge of the velocity of the particle ($+v_P$), which contradicts the essence of non-relativistic quantum mechanics when the Hamiltonian is such that the velocity operator is proportional to the momentum: $\mathbf{v} = \frac{d\mathbf{r}}{dt} = \frac{i}{\hbar}[\hat{H}, \mathbf{r}] = \frac{\mathbf{p}}{m_0}$. This is the case of our microscopic Hamiltonian, provided the contribution of the spin-orbit coupling term to the velocity operator is small (in such a case $\hat{H} \approx \frac{\hat{p}^2}{2m_0} + \hat{V}$). When $\mathbf{v} = \frac{\mathbf{p}}{m_0}$, the velocity and the position ($\mathbf{r}$) operators cannot be simultaneously known with arbitrary accuracy, as follows from the Heisenberg uncertainty relations for **p** and **r**. Because for the one-dimensional problem **v** can only assume two values in case of a zero-gap material ($\pm v_P$), the knowledge of the direction of propagation (and thus of the velocity with no uncertainty) precludes *any* type of localization of the wave packet. The solution (60) is consistent with such a fundamental restriction. This also highlights that the effective medium Schrödinger equation (56) is fundamentally different from the wave equation (58), because the latter admits solutions that enable both the localization of the position and of the velocity of the wave packet, whereas the former does not. It is important to emphasize that the effective medium Hamiltonian only enables characterizing the *time evolution* of "macroscopic states", and thus many relevant initial time states (for which the time evolution is not consistent with Eq. (60)) are out of reach of the effective medium description. The most notorious example is the energy eigenstates, which evidently *do not* vary in time as predicted by Eq. (60). We remind that for a zero-gap semiconductor the energy eigenstates are not "macroscopic states". The restrictions on the use of the effective medium Hamiltonian in the calculation of the electronic band structure are less severe than in the time evolution problem, and, as discussed previously, it holds enough information to calculate exactly the energy dispersion of the stationary states.

To conclude, we would like to note that the time evolution of the initial state $\Psi_{t=0}(x)$ does not preserve the norm $\int d^3\mathbf{r} |\Psi|^2$. Obviously, at the "microscopic level" the time evolution of the exact wavefunction $\psi$ preserves the norm $\int d^3\mathbf{r} |\psi|^2$. The reason for the discrepant behavior was already mentioned in Ref. [30], and is related to the fact that $\Psi = \{\psi\}_{av}$ does not imply that $|\Psi|^2 = \{|\psi|^2\}_{av}$, and hence in general $|\Psi|^2$ does *not* correspond to the spatially-averaged probability density. It will be proven elsewhere that for stationary states the spatially-averaged probability density can be written in terms of $\Psi$ and of the effective medium Hamiltonian.

## VIII. Conclusion

Using the effective medium approach derived in our previous work [30], we calculated from "first-principles" the effective Hamiltonian of a bulk material. Within the eight-band Kane approximation, the effective Hamiltonian for bulk semiconductor compounds with a zincblende structure can be calculated explicitly and is a scalar operator given by Eq. (29). For excitations associated with energies close to the edges of either the conduction or valence bands, the effective Hamiltonian reduces to the simpler form $H_{ef}(-i\nabla, E) \approx -\frac{\hbar^2}{2}\nabla\left(\frac{1}{M_{ef}}\nabla\right) + V_{ef}(E)$, which is consistent with the formalism of G. Bastard, i.e. with the envelope function approximation. Our results highlight that the envelope function approximation is related to the effective medium theory used in the context of electromagnetic metamaterials. Using the developed theory, we discussed the electronic structure of several bulk semiconductor compounds, emphasizing the analogies with electromagnetic metamaterials. Finally, we discussed the time evolution of a macroscopic electron wave in a zero-gap semiconductor with a linear energy-momentum relation, and that the dynamics of the effective medium Hamiltonian is consistent with the uncertainty relations.

**Acknowledgments:**

This work is supported in part by the U.S. Air Force of Scientific Research (AFOSR) grant numbers FA9550-08-1-0220 and FA9550-10-1-0408, and by Fundação para a Ciência e a Tecnologia grant number PTDC/EEATEL/100245/2008.

# Appendix

Here, we derive the relation between the coefficients $c_E$ and $c_H$ in the expansion $\psi_{t=0} = e^{i\mathbf{k}\cdot\mathbf{r}} \approx c_E \psi_{c,\mathbf{k}}(\mathbf{r}) + c_H \psi_{v,\mathbf{k}}(\mathbf{r})$ considered in Sect. VII. Using the ket notation, we denote $|E_1\mathbf{k}\rangle$ and $|H_1\mathbf{k}\rangle$ the states associated with $\psi_{c,\mathbf{k}}(\mathbf{r})$ and $\psi_{v,\mathbf{k}}(\mathbf{r})$, respectively (the spin quantum number is omitted in some formulas for simplicity). From Sect. IV, it is clear that $|E_1\mathbf{k}\rangle$ and $|H_1\mathbf{k}\rangle$ can be written in terms of the kets $|S\sigma\rangle$, $|X\sigma\rangle$, $|Y\sigma\rangle$ and $|Z\sigma\rangle$. In case the split-off energy $\Delta$ is much larger than the band-gap energy, it is possible to write [see Eq. (14) and (17) of Ref. [21]; only one of the degenerate states is shown]:

$$\mathbf{T}_{-\mathbf{k}}|E_1\mathbf{k}\rangle = a_c |iS\downarrow\rangle + b_c \left|\frac{1}{\sqrt{2}}(X-iY)\uparrow\right\rangle + c_c |Z\downarrow\rangle, \tag{A1}$$

$$\mathbf{T}_{-\mathbf{k}}|H_1\mathbf{k}\rangle = a_v |iS\downarrow\rangle + b_v \left|\frac{1}{\sqrt{2}}(X-iY)\uparrow\right\rangle + c_v |Z\downarrow\rangle. \tag{A2}$$

where

$$a_c = \left(\frac{\eta + E_g}{2\eta}\right)^{1/2}, \qquad b_c = \left(\frac{\eta - E_g}{6\eta}\right)^{1/2}, \qquad c_c = \left(\frac{\eta - E_g}{3\eta}\right)^{1/2}, \tag{A3}$$

$$a_v = -\left(\frac{\eta - E_g}{2\eta}\right)^{1/2}, \qquad b_v = \left(\frac{\eta + E_g}{6\eta}\right)^{1/2}, \qquad c_v = \left(\frac{\eta + E_g}{3\eta}\right)^{1/2} \tag{A4}$$

and $\eta$ is a (*k*-dependent) parameter defined in Ref. [21]. The operator $\mathbf{T}_{-\mathbf{k}}$ is such that $\langle \mathbf{r}|\mathbf{T}_{-\mathbf{k}}|\theta\rangle = e^{-i\mathbf{k}\cdot\mathbf{r}}\theta(\mathbf{r})$ with $\theta(\mathbf{r}) = \langle \mathbf{r}|\theta\rangle$.

The important point is that if the band gap energy $E_g$ vanishes (a zero-gap semiconductor), we have $a_c = -a_v = 1/\sqrt{2}$, $b_c = b_v = 1/\sqrt{6}$, and $c_c = c_v = 1/\sqrt{3}$. Thus, the initial time macroscopic state $|\psi\rangle = c_E|E_1\mathbf{k}\rangle + c_H|H_1\mathbf{k}\rangle$ is such that:

$$\mathbf{T}_{-\mathbf{k}}|\psi\rangle \approx (c_E - c_H)\frac{1}{\sqrt{2}}|iS\downarrow\rangle + (c_E + c_H)\frac{1}{\sqrt{6}}\left|\frac{1}{\sqrt{2}}(X-iY)\uparrow\right\rangle + (c_E + c_H)\frac{1}{\sqrt{3}}|Z\downarrow\rangle. \quad (A5)$$

Since $\psi(\mathbf{r}) = e^{i\mathbf{k}\cdot\mathbf{r}}$ if we calculate the scalar product of both sides of the equation with a ket of the form $\left|\frac{1}{\sqrt{2}}(X-iY)\uparrow\right\rangle$ (or alternatively $|Z\downarrow\rangle$) it is immediately found that $(c_E + c_H) = 0$. This is so because the kets $|X\sigma\rangle$, $|Y\sigma\rangle$ and $|Z\sigma\rangle$ have the same symmetry as *p*-type orbitals, and hence $\langle m|\mathbf{T}_{-\mathbf{k}}|\psi\rangle = 0$ with *m=X,Y,Z*. Thus it follows that $c_E = -c_H$, which gives the desired relation between the two coefficients.

We also note that because the result of spatial-averaging the kets $|X\sigma\rangle$, $|Y\sigma\rangle$ and $|Z\sigma\rangle$ is zero [Eq. (23)] one can write $\Psi_{c,\mathbf{k}}e^{i\mathbf{k}\cdot\mathbf{r}} = \{\psi_{c,\mathbf{k}}\}_{av} = ia_c e^{i\mathbf{k}\cdot\mathbf{r}}u_{s,av}$ where $u_{s,av}$ is some non-zero constant, defined as in Eq. (14) for *m=S*. Similarly, it is easy to check that $\Psi_{v,\mathbf{k}}e^{i\mathbf{k}\cdot\mathbf{r}} = \{\psi_{v,\mathbf{k}}\}_{av} = ia_v e^{i\mathbf{k}\cdot\mathbf{r}}u_{s,av}$. Since for a zero-gap semiconductor $a_c = -a_v$ we finally conclude that $\Psi_{c,\mathbf{k}} = -\Psi_{v,\mathbf{k}}$.

## References

[1] J. D. Joannopoulos, S. G. Johnson, J. N. Winn, and R. D. Meade, *Photonic Crystals: Molding the Flow of Light*, Princeton University Press, Princeton NJ, 2008.